\documentclass{aastex}

\shorttitle{MOA-2011-BLG-274}
\shortauthors{Matthew Freeman et al.}

\usepackage{graphicx}
\usepackage{epstopdf}
\usepackage{amsmath}
\usepackage{natbib}

\begin{document}

\title{Can the masses of isolated planetary-mass gravitational lenses be measured by terrestrial parallax?}
\author{M.~Freeman$^*$\altaffilmark{1}, L.C.~Philpott\altaffilmark{2}, F.~Abe\altaffilmark{3}, M.D.~Albrow\altaffilmark{4}, D.P.~Bennett\altaffilmark{5}, I.A.~Bond\altaffilmark{6},C.S.~Botzler\altaffilmark{1}, J.C.~Bray\altaffilmark{1},  J.M.~Cherrie\altaffilmark{1}, G.W.~Christie\altaffilmark{7}, Z.~Dionnet\altaffilmark{8}, A.~Gould\altaffilmark{9}, C.~Han\altaffilmark{10}, D.~Heyrovsk$\acute{\mathrm{y}}^{11}$, J.M.~McCormick\altaffilmark{12},  D.M.~Moorhouse\altaffilmark{13}, Y.~Muraki\altaffilmark{3},T.~Natusch\altaffilmark{7}, , N.J.~Rattenbury\altaffilmark{1}, J.~Skowron\altaffilmark{14}, T.~Sumi \altaffilmark{15}, D.~Suzuki\altaffilmark{15}, T.-G.~Tan\altaffilmark{16}, P.J.~Tristram\altaffilmark{17} and P.C.M.~Yock\altaffilmark{1}}
 
\email{$^*$mfre070@aucklanduni.ac.nz}

\altaffiltext{1}{Department of Physics, University of Auckland, Private Bag 92019, Auckland 1142, New Zealand}
\altaffiltext{2}{Department of Earth, Ocean and Atmospheric Sciences, University of British Columbia, Vancouver, British Columbia, V6T 1Z4, Canada}
\altaffiltext{3}{Solar-Terrestrial Environment Laboratory, Nagoya University, Nagoya 464-8601, Japan}
\altaffiltext{4}{Department of Physics and Astronomy, University of Canterbury, P.O.~Box 4800, Christchurch 8020, New Zealand}  
\altaffiltext{5}{Department of Physics, 225 Nieuwland Science Hall, University of Notre Dame, Notre Dame, IN 46556, USA}
\altaffiltext{6}{Institute for Information and Mathematical Sciences, Massey University, Private Bag 102-904, Auckland 1330, New Zealand}
\altaffiltext{7}{Auckland Observatory, PO Box 180, Royal Oak, Auckland 1345, New Zealand}
\altaffiltext{8}{Universit\'{e} d'Orsay, bat 470, 91400 Orsay, France} 
\altaffiltext{9}{Department of Astronomy, Ohio State University, 140 West 18th Avenue, Columbus, OH 43210, USA}
\altaffiltext{10}{Department of Physics, Chungbuk National University, 410 Seongbong-Rho, Hungduk-Gu, Chongju 371-763, Korea}
\altaffiltext{11}{Institute of Theoretical Physics, Charles University in Prague, Faculty of Mathematics and Physics, V Holesovickach 2, 18000 ~~Prague, Czech Republic}
\altaffiltext{12}{Farm Cove Observatory, 2/24 Rapallo Place, Pakuranga, Auckland 2012, New Zealand}
\altaffiltext{13}{Kumeu Observatory, Kumeu, New Zealand}
\altaffiltext{14}{Warsaw University Observatory, Al. Ujazdowskie 4, 00-478, Warszawa, Poland}
\altaffiltext{15}{Department of Earth and Space Science, Osaka University, 1-1 Machikaneyama-cho, Toyonaka, Osaka 560-0043, Japan} 
\altaffiltext{16}{Perth Exoplanet Survey Telescope, Perth, Australia}
\altaffiltext{17}{Mt.~John University Observatory, P.O.~Box 56, Lake Tekapo 8770, New Zealand}

\begin{abstract}
Recently Sumi et al.~(2011) reported evidence for a large population of planetary-mass objects (PMOs) that are either unbound or orbit host stars in orbits $\geq10$ AU. Their result was deduced from the statistical distribution of durations of gravitational microlensing events observed by the MOA collaboration during 2006 and 2007. Here we study the feasibility of measuring the mass of an individual PMO through microlensing by examining a particular event, MOA-2011-BLG-274. This event was unusual as the duration was short, the magnification high, the source-size effect large and the angular Einstein radius small. Also, it was intensively monitored from widely separated locations under clear skies at low air masses. Choi et al.~(2012) concluded that the lens of the event may have been a PMO but they did not attempt a measurement of its mass. We report here a re-analysis of the event using re-reduced data. We confirm the results of Choi et al.~ and attempt a measurement of the mass and distance of the lens using the terrestrial parallax effect. Evidence for terrestrial parallax is found at a $3\sigma$ level of confidence. The best fit to the data yields the mass and distance of the lens as $0.80\pm$ 0.30 $M_{\mathrm{J}}$ and $0.80\pm0.25$ kpc respectively. We exclude a host star to the lens out to a separation $\sim$ 40 AU. Drawing on our analysis of MOA-2011-BLG-274 we propose observational strategies for future microlensing surveys to yield sharper results on PMOs including those down to super-Earth mass.
\end{abstract}


\section{Introduction}
Three avenues of research have yielded evidence for isolated planetary-mass objects (PMOs) in recent years. Observations of star-forming regions by several groups indicate the presence of isolated objects with masses of a few Jupiter masses (Barrado-y-Navascues et al.~2001, Burgess et al.~2009, Haisch et al.~2010, Marsh et al.~2010, Delorme et al.~2012, Pena Ramirez et al.~2012, Scholz et al.~2012, Beichman et al.~2013, Liu et al.~2013). 

Evidence has also been reported by gravitational microlensing for objects having approximately the mass of Jupiter distributed throughout the Galaxy (Sumi et al.~2011). A surprisingly large number of these objects were inferred, nearly two for every star in the Galaxy. However, the evidence was based on the statistical properties of a large sample of microlensing events. Measurements of the masses of individual PMOs were not attempted. 

Most recently, Luhman (2014) reported evidence for a nearby ($\sim$2pc) Y type brown dwarf with temperature $\sim$250K, mass 3-10$M_\mathrm{J}$ and age 1-10Gyr. The search by Luhman was aimed specifically at nearby brown dwarfs, and the result suggests a high spatial density of them, comparable to that reported by Sumi et al.~(2011) for isolated PMOs with masses $\sim\mathrm{M}_\mathrm{J}$. It is possible that the object found by Luhman represents the high mass tail of the distribution found by Sumi et al. Further observations of nearby objects may therefore help to identify the objects found by Sumi et al.      

The above results, in particular those by microlensing, raise a number of questions. Did these objects form in-situ, and could they therefore be better classified as sub-brown-dwarfs (Gahm et al.~2013)? Or could they be planets orbiting stars at larger radii than gravitational microlensing is sensitive to (Quanz et al.~2012)? Were they planets that formed around stars, and subsequently underwent ejection through planet-planet or planet-star interactions (Guillochon et al.~2011, Malmberg et al.~2011, Veras \& Raymond 2012, Kaib et al.~2013)? Is the high number of putative objects of Jupiter mass accompanied by a larger population of terrestrial mass objects? Or does the high number of apparent detections imply some degree of bias in the statistical procedure used, or contamination by variable stars in the sample masquerading as gravitational microlensing events with low-mass lenses? 
 
In this paper we report a re-analysis of MOA-2011-BLG-274, a microlensing event in which the lens was reported by Choi et al.~(2012) as a possible PMO. 

Our notation is as follows: The parameters $r_{\mathrm{E}}$, $t_{\mathrm{E}}$, and $t_{0}$ denote the Einstein radius, crossing time and time of closest approach between the lens and source stars, and $\theta_{\mathrm{E}}$, $\theta_{\mathrm{min}}$, and $\theta_{\mathrm{S}}$ denote the angular Einstein radius, the impact parameter between the lens and source stars in angular coordinates, and the angular radius of the source star respectively. $D_\mathrm{L}$ and $D_\mathrm{S}$ denote the distances to the lens and source stars. We define $u_{\mathrm{min}}=\theta_{\mathrm{min}}/\theta_{\mathrm{E}}$ and ${\rho}={\theta_{\mathrm{S}}}/{\theta_{\mathrm{E}}}$. Finally, for planetary events, we denote by $q$, $d$, and $\psi$ the planet:star mass ratio, projected separation in units of $r_{\mathrm{E}}$ and axis relative to the source star track respectively. The fundamental parameter of microlensing, the Einstein radius $r_{\mathrm{E}}$, is defined in Liebes~(1964).

Choi et al.~reported analyses of several microlensing events where the lens geometrically transited the source, i.e. where $\theta_{\mathrm{min}}\leq\theta_\mathrm{S}$ or, equivalently, $u_{\mathrm{min}}\leq\rho$. They referred to these well-aligned events as ones in which the lens `passed over' the source star. In such events it is possible to measure the angular Einstein radius $\theta_{\mathrm{E}}$ of the event (Gould 1994, Alcock et al.~1997). Choi et al.~found $\theta_{\mathrm{E}}$ $\sim$ 0.08 mas for MOA-2011-BLG-274, which is unusually small in comparison to typical microlensing events. They also reported that the Einstein radius crossing time, $t_{\mathrm{E}}$, for the event was unusually small, $\sim$ 2.7 days. As the Einstein radius for gravitational lensing is proportional to the square root of the mass of the lens, Choi et al.~concluded that the lens of the event, i.e. MOA-2011-BLG-274L, could have been a PMO.  

Here we extend the analysis of Choi et al.~in a number of ways. The photometry for two of the datasets was improved by re-reductions. Secondly, instrument-specific limb-darkening coefficients of the source star  were computed and used in the analysis. Thirdly, an investigation into the possibility of recovering the lens mass and distance through measurement of terrestrial parallax was made following a prediction of Gould (1997). Fourthly, an independent search for satellites orbiting the putative PMO was carried out and the results compared to those of Choi et al. Fifthly, a search was made for a star in the vicinity of MOA-2011-BLG-274L that it might be orbiting. Finally, we discuss future observational strategies that could yield definitive measurements of the masses of individual PMOs and also uncover PMOs of lower mass than those reported by Sumi et al. Our analysis utilised code based on magnification maps (Abe et al.~2013) that was written independently of the code used by Choi et al.   

\section{MOA-2011-BLG-274 data}
MOA-2011-BLG-274 was discovered and alerted on 2011 June 29 at 9:44:27 UT as a possible microlensing event of high magnification at RA $17^h 54^m 42^s.34$ and declination $-28^\circ54'59~26"$, or (\textit{l,b}) = (1.04$^\circ$, -1.70$^\circ$), as part of the nightly survey of the Galactic bulge that has been conducted by the MOA collaboration during southern winters since 2000 from the Mt John University Observatory in New Zealand (Bond et al.~2001)\footnote{http://www.phys.canterbury.ac.nz/moa/}. 

The MicroFUN collaboration\footnote{http://www.astronomy.ohio-state.edu/$\sim$microfun/} responded promptly to the alert. They observed MOA-2011-BLG-274 intensively over a three hour period when the magnification rose from $\sim$ 40 to a peak value of $\sim$ 200 and then fell again to $\sim$ 40. The peak occurred over Australasia and MicroFUN monitored this period with a 0.4m telescope at the Auckland Observatory equipped with a custom red filter, a 0.36m telescope at Kumeu (in the Auckland region) equipped with a similar red filter, a 0.36m unfiltered telescope at the Farm Cove Observatory in Auckland, and a 0.3m unfiltered telescope at the PEST observatory in Perth.

\begin{figure}
\epsscale{1.0}
\plotone{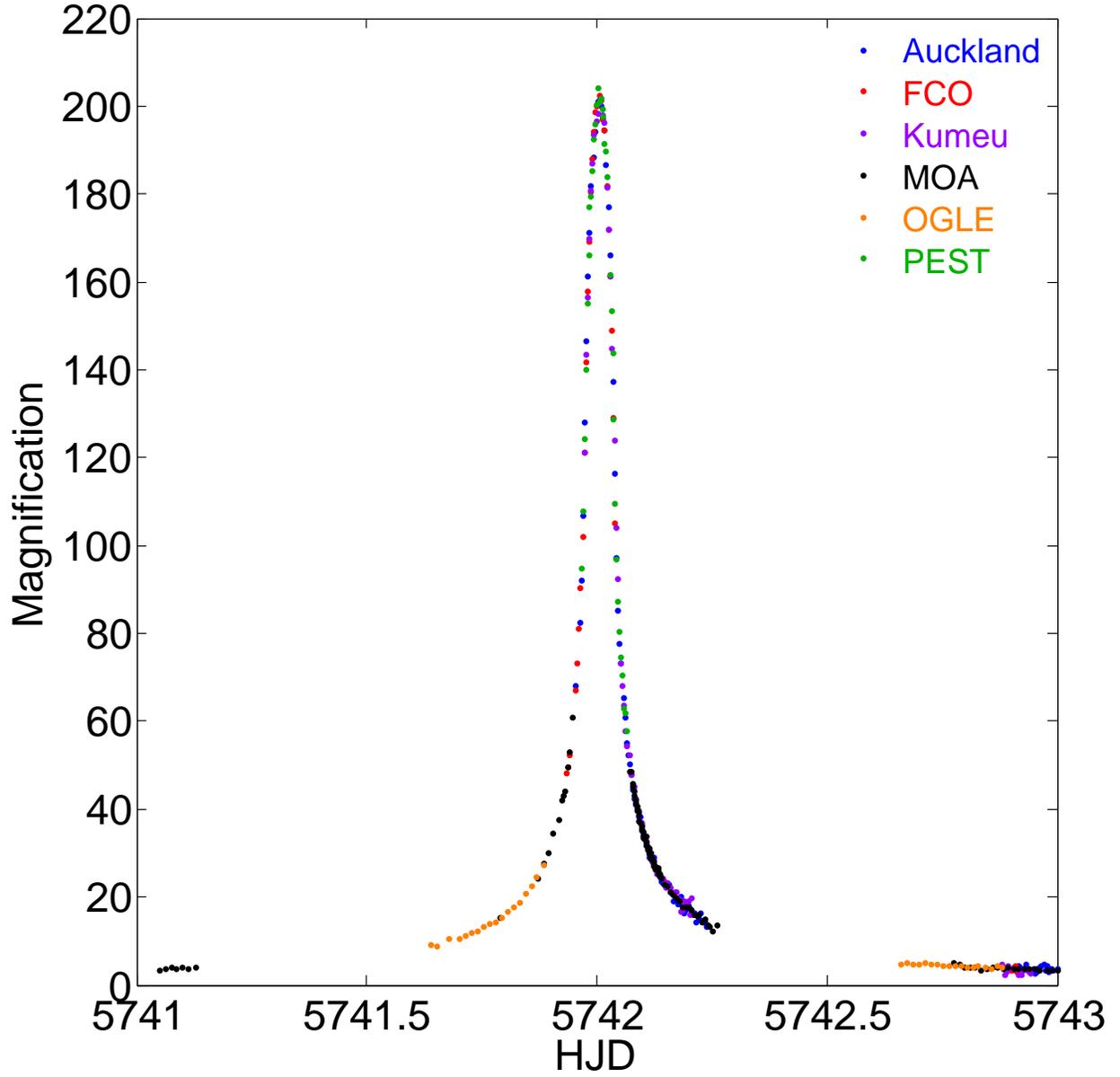}
\caption{Two day segment of the lightcurve of MOA-2011-BLG-274 showing the datasets from the MOA and OGLE survey telescopes, and data provided by MicroFUN telescopes at Auckland and Perth.}
\label{fig1}
\end{figure}

Unfortunately, conditions were cloudy at Mt John during the peak of the event, although the sky was clear immediately prior to and immediately following the peak when the magnification was $\sim40$. Data were recorded with the 1.8m MOA telescope and the custom MOA red filter at these times, and also on previous and subsequent nights. In addition, data were obtained at magnifications $\sim $10 and lower in the I-band by the OGLE\footnote{http://ogle.astrouw.edu.pl/} and CTIO\footnote{http://www.astro.yale.edu/smarts/index.htm} 1.3m telescopes in Chile. The larger datasets are shown in Figure~\ref{fig1}. 

To reduce computer runtime, the MOA data were binned into single night segments in the wings of the event. Also, to minimise the effects of possible drifts in the MOA data, all data taken prior to 2010 were excluded. 

The data were generally reduced by the groups that supplied them. Thus the MOA and OGLE data were reduced using the codes of Bond et al.~(2001) and Udalski (2003). These codes utilise the difference imaging procedure of Alard and Lupton~(1998), and Wozniak~(2000). In the case of the MOA data the online reduction was replaced by an off-line reduction to optimise the precision through improved astrometry. The MicroFUN data were reduced by the DoPHOT code of Schechter et al.~(1993) with the exception of the PEST data for which the latter reduction appeared relatively noisy. These data were therefore re-reduced using the more accurate difference imaging pySIS code of Albrow et al.~(2009). 

The uncertainties in the data from the Auckland Observatory, PEST and MOA telescopes were renormalised to force their contributions to $\chi^2$ to be approximately equal to the numbers of data points. For the MOA data this involved adding uncertainties in quadrature and scaling them. The data from the other telescopes were already reasonably well normalised. A total of 1276 data points were used in the final analysis with 683 by MOA, 335 by OGLE, 113 by Auckland, 35 by Farm Cove, 77 by Kumeu, and 33 by PEST.

\section{Source star} 
It is customary to identify the source star in a gravitational microlensing event of high magnification from its de-reddened colour $(V-I)_{s,0}$ and its de-reddened apparent magnitude $I_{s,0}$. These are normally determined by recording a few images of the event at high magnification in the V passband. These are combined with contemporaneous or near-contemporaneous images in the I passband to yield an instrumental and reddened $(V-I)_\mathrm{s,i}$ colour and an instrumental and reddened apparent magnitude $I_\mathrm{s,i}$ of the source star. These are then converted to non-instrumental and dereddened values $(V-I)_{s,0}$ and $I_{s,0}$ by using the position of the red clump on the colour magnitude diagram as a standard. 
    
No V band images of MOA-2011-BLG-274 were taken at high magnification. A modified version by Gould et al.~(2010) of the above procedure was therefore used. Instrumental magnitudes of the source star were obtained from single lens fits to the light curves recorded by the unfiltered PEST telescope, $R_\mathrm{s,p,i}$, and the OGLE telescope, $I_\mathrm{s,o,i}$. From these an instrumental colour $R_\mathrm{s,p,i}$ - $I_\mathrm{s,o,i}$ = 2.13 $\pm$ 0.04 was deduced. 

Using field stars, two linear colour-colour relationships were found: one for $R_\mathrm{p,i}$-$I_\mathrm{c,i}$ vs $V_\mathrm{c,i}$-$I_\mathrm{c,i}$ and one for $I_\mathrm{o,i}$-$I_\mathrm{c,i}$ vs $V_\mathrm{c,i}$-$I_\mathrm{c,i}$, where $V_\mathrm{c,i}$ and $I_\mathrm{c,i}$ denote instrumental magnitudes recorded by the CTIO telescope in the V and I passbands. These two relationships were subtracted to yield a linear relationship between $R_\mathrm{p,i}$-$I_\mathrm{o,i}$ and $V_\mathrm{c,i}$-$I_\mathrm{c,i}$. Insertion of the instrumental colour $R_\mathrm{s,p,i}$ - $I_\mathrm{s,o,i}$ into this relationship yielded the instrumental colour of the source star seen by the CTIO telescope $V_{s,c,i}-I_{s,c,i}$ = 0.0 $\pm$ 0.08. Finally, this was converted to a dereddened and non-instrumental colour $(V-I)_{s,0}$ by reference to the position of the red clump on a colour magnitude diagram seen by the CTIO telescope in the normal manner. This procedure also yielded the de-reddened and non-instrumental magnitude of the source star $I_{s,0}$. The final results were  

\begin{equation}
(V-I)_{s,0}  =  0.76 \pm 0.10,
\end{equation}
\begin{equation}  
I_{s,0}  = 17.96 \pm 0.10.
\end{equation}

The angular radius $\theta_\mathrm{S}$ of the source star was determined from the above results and the surface brightness relationship of Kervella (2008)\footnote{This combines the inverse square and Stefan Boltzmann laws via an empirical correction for stellar spectra.}. This yielded  
\begin{equation}
\theta_{\mathrm{S}} = 0.87\pm0.12~\mu\mathrm{as} .
\end{equation}

Choi et al.~followed similar procedures to those described above in their analysis of MOA-2011-BLG-274 and obtained results within $1\sigma$ of those above. 

The distance to the source star was determined under the assumption that it lay in the Galactic bar. Adopting the model of Cao et al.~(2013) of the Galactic bar we deduce a distance to the source for MOA-2011-BLG-274 as  

\begin{equation}
D_{\mathrm{S}}=7.9\pm0.6~\mathrm{kpc}.
\end{equation}

Equation (3) yields the source star radius $r_\mathrm{S}$ and absolute magnitude $M_\mathrm{I}$ as     

\begin{equation}
r_{\mathrm{S}}=1.47\pm0.24 ~r_{\mathrm{solar}}, 
\end{equation}
\begin{equation}
M_{\mathrm{I}}=3.47\pm0.20.
\end{equation}
   
The above results enable the source star to be classified. Recently Bensby et al.~(2011) used a sample of high-magnification microlensing events to spectroscopically determine the distribution of ages and metalicities of main sequence stars in the Galactic bulge. A roughly bimodal distribution was found containing approximately equal mixtures of older metal-poor stars with ages 8-12 Gy and metalicities [Fe/H] $\sim$ -0.6, and younger metal-rich stars with ages 4-12 Gy and metalicities [Fe/H] $\approx$ +0.3. A subsequent investigation with a larger sample of stars indicated additional stars with intermediate metalicities (Bensby et al.~2013).

\begin{figure}
\plotone{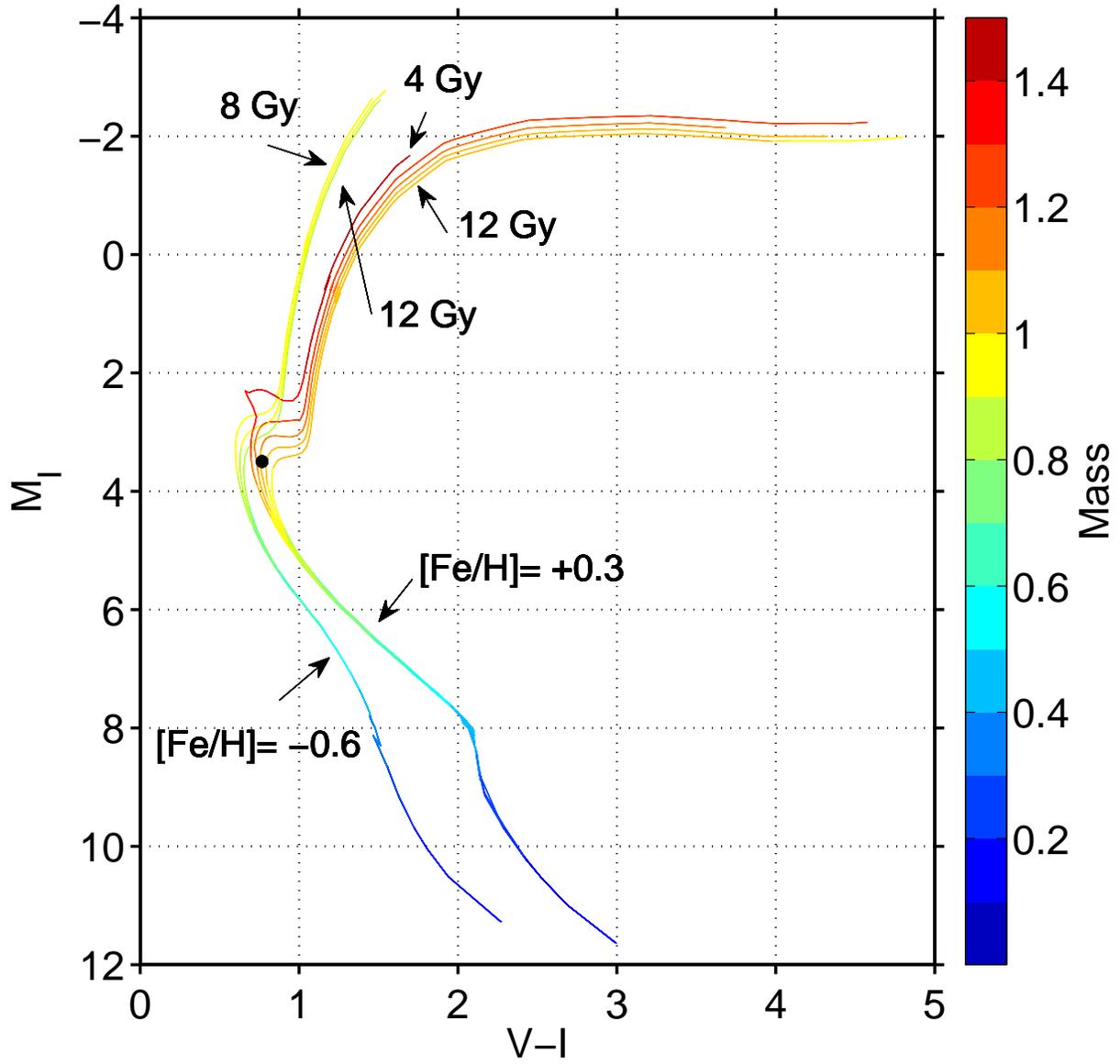}
\caption{Dependence of $M_I$ on $V-I$ as determined by the isochrones of Girardi et al.~(2002) for stars of various ages and [Fe/H] = -0.6 or +0.3. The source star for MOA-2011-BLG-274 is indicated by the black circle.} 
\label{fig2}
\end{figure}

\begin{figure}
\plotone{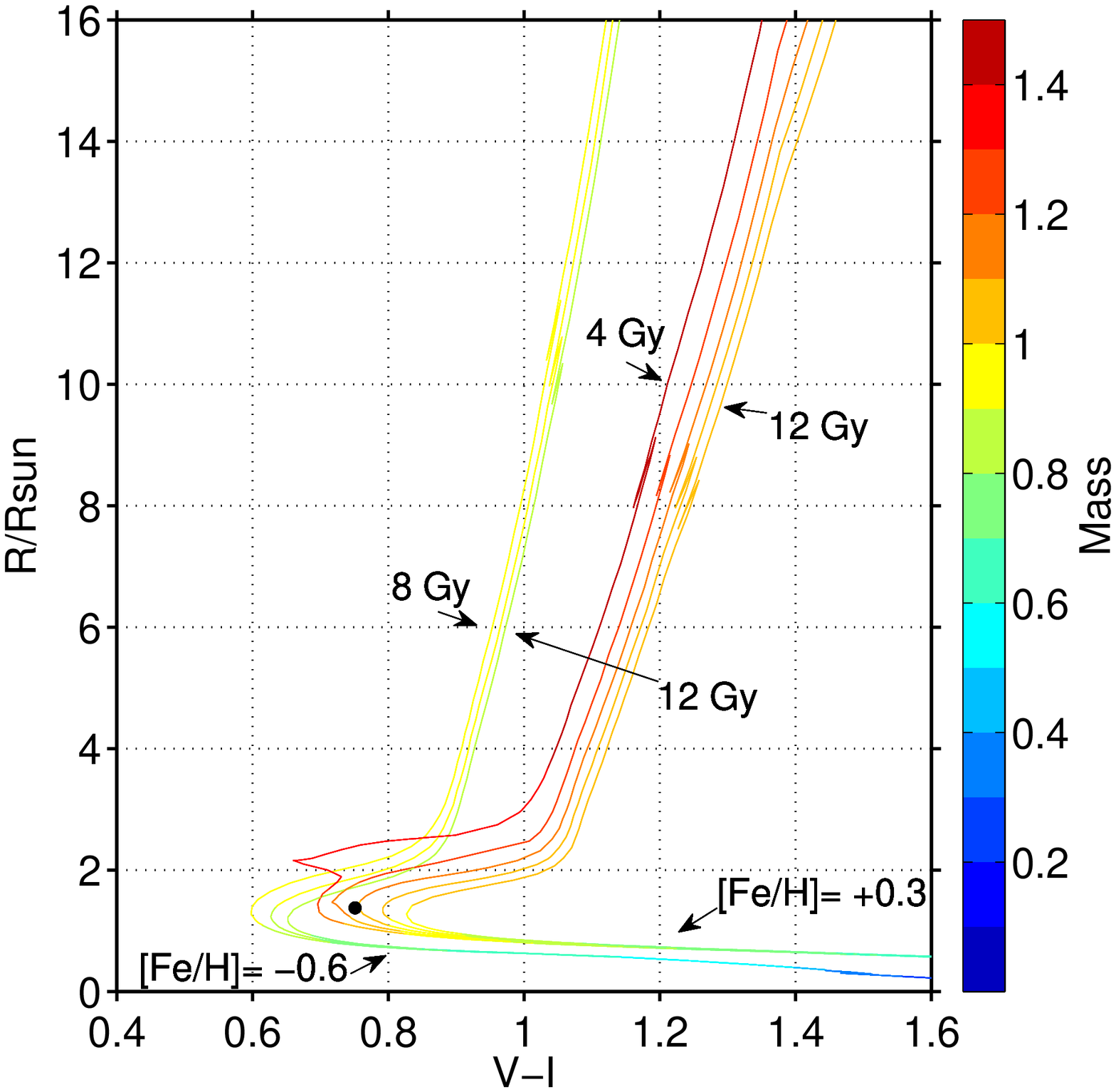}
\caption{Dependence of stellar radius on $V-I$ as determined by the isochrones of Girardi et al.~(2002) for stars of various ages and [Fe/H] = -0.6 or +0.3. The source star for MOA-2011-BLG-274 is indicated by the black circle.} 
\label{fig3}
\end{figure}

The dependence of stellar radius and absolute magnitude $M_{\mathrm{I}}$ on colour $(V-I)$ for stars in the above ranges may be determined from the isochrones of Girardi et al.~(2002) as shown in Figures~\ref{fig2} and~\ref{fig3}. The above values of source star radius $r_{\mathrm{S}}$ and absolute magnitude $M_{\mathrm{I}}$ for MOA-2011-BLG-274 then enable the source star to be identified as most likely an old, metal-rich, turn-off star of $\sim$ 1.1 solar mass, as shown on Figures~\ref{fig2} and~\ref{fig3}. These plots should be applicable to most microlensing events.

An effective temperature for the source star of $5700\pm200$K was estimated from its colour (Bessell et al, 1998). In comparison, Choi et al.~(2012) reported a temperature of 6000K in their analysis. Limb-darkening coefficients were obtained from Kurucz's ATLAS9 stellar atmosphere models using the method described by Heyrovsky (2007). In Table~\ref{table1} we present their values computed specifically for each light curve, taking into account the filter transmission, the CCD quantum efficiency, as well as the extinction towards the Galactic-bulge source. We point out that the values are in good agreement with those of Choi et al.~(2012). They quoted measurements and uncertainties for the Auckland and Kumeu data only. These are within 1 and 2 standard deviations of ours.  

\begin{deluxetable}{l c c c}
\tablewidth{0pt}
\tablecaption{Limb darkening coefficients of the source star of MOA-2011-BLG-274.\label{table1}}
\tablehead{\colhead{Telescope} & \colhead{Linear} & \multicolumn{2}{c}{Square-root}\\
& \colhead{u} & \colhead{c} & \colhead{d}}
\startdata
Auckland & 0.5194 & 0.1255 & 0.6119 \\
FCO & 0.5374 & 0.1549 & 0.5943 \\
Kumeu & 0.5187 & 0.1248 & 0.6117 \\
MOA & 0.5137 & 0.1142 & 0.6205 \\
OGLE & 0.4801 & 0.0829 & 0.6164 \\
PEST & 0.5374 & 0.1549 & 0.5943 \\ 
\enddata
\tablecomments{We assume $A_I=1.8$, $A_V=3.0$ (Sumi et al., 2003), temperature = 5700K, log g = 4.15, $[Fe/H]=+0.3$ and $v_t=2.0$ km/s.}
\end{deluxetable}

\section{Light curve without parallax}

An attempt was initially made to model the data on MOA-2011-BLG-274 in a point source approximation but this proved to be entirely impossible. Very large departures were found that clearly showed the lens in the event had transited the source star.   

\begin{figure}
\plotone{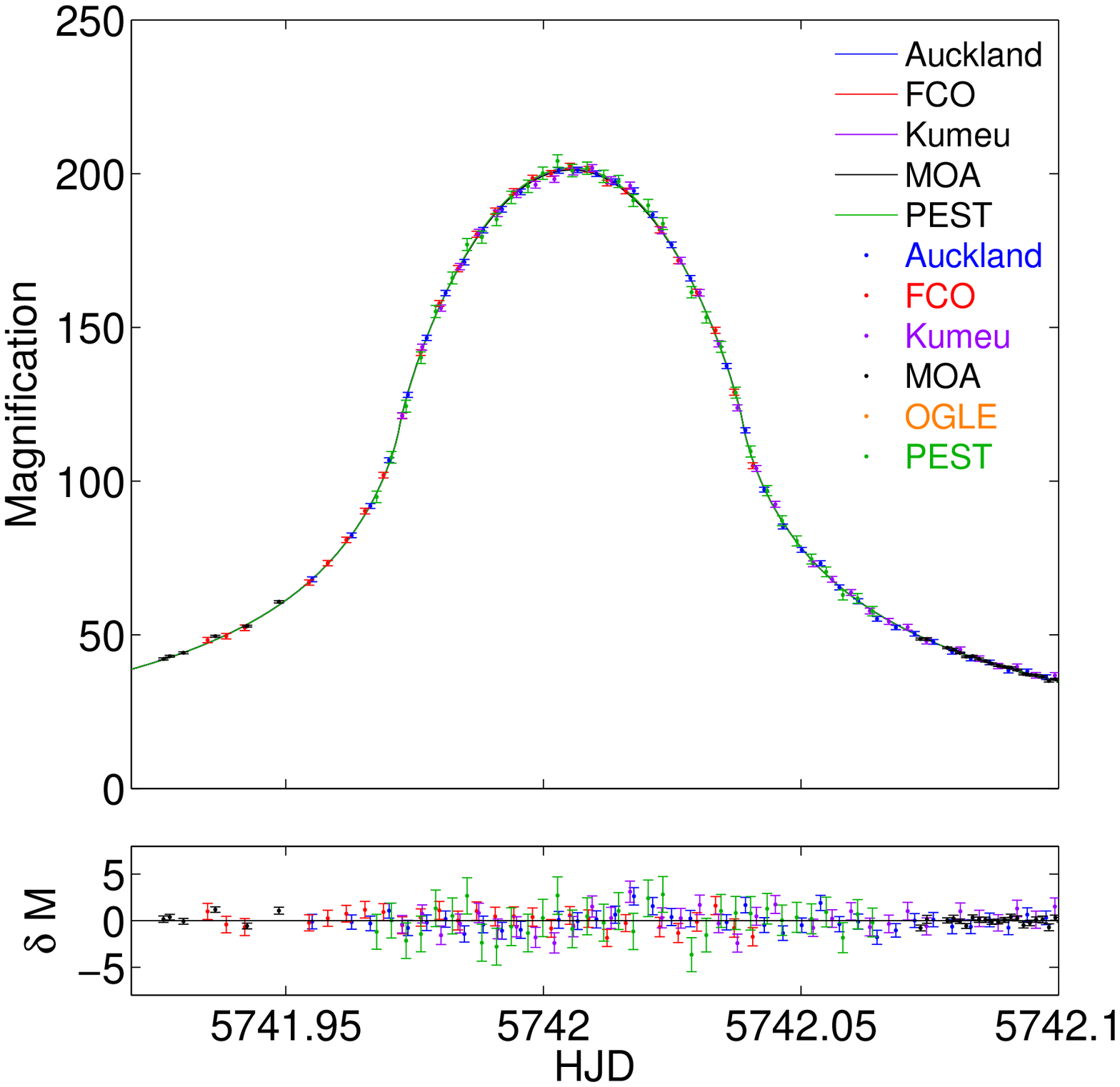}
\caption{Best fit light curves for the individual telescopes assuming a finite, linearly limb-darkened source at 5700K but excluding parallax. Residuals shown below}
\label{fig4}
\end{figure}

Figure~\ref{fig4} shows the best fit to the data assuming a finite, linearly limb-darkened source at 5700K, with limb-darkening coefficients given in Table~\ref{table1}, but with no allowance made for parallax. The parameters for this fit, and for others with source temperatures of 5500K and 5900K and square-root limb darkening, are given in Table~\ref{table2}. These are compared to earlier results obtained by Choi et al.~(2012). The largest difference occurs in the value of $t_\mathrm{E}$ which is to be expected following the re-reduction of the baseline data by MOA. 

\begin{deluxetable}{c c c c c c c c c}
\tabletypesize{\scriptsize}
\tablewidth{0pt}
\tablecaption{Best fitting parameters without parallax.\label{table2}}
\tablehead{
\colhead{Source} & \colhead{limb darkening} & \colhead{$u_\mathrm{min}$} & \colhead{$\rho$} & \colhead{$t_0$ (HJD)} & \colhead{$t_{\mathrm{E}}$ (d)} & \colhead{$\theta_{\mathrm{E}}$ (mas)} & \colhead{$\mu$ (mas y$^{-1}$)} & \colhead{$\chi^2$}
}
\startdata
Present work & linear (5700K) & 0.00228 & 0.0105 & 2545742.00547 & 3.26 & 0.083 & 9.29 & 1349.5\\
Present work & linear (5500K) & 0.00224 & 0.0104 & 2545742.00547 & 3.30 & 0.084 & 9.29 & 1350.0\\
Present work & linear (5900K) & 0.00232 & 0.0105 & 2545742.00546 & 3.25 & 0.083 & 9.32 & 1350.3\\
Present work & square root (5700K) & 0.00228 & 0.0105 & 2545742.00547 & 3.26 & 0.083 & 9.29 & 1350.6\\
Choi et al.~(2012) & linear & 0.0029 & 0.0129 & 2545742.005 & 2.65 & 0.08 & 11.18 & -\\ 
\enddata
\tablecomments{{Linear} limb darkening coefficients for a source star temperature of 5500K or 5900K were computed keeping the other atmosphere parameters fixed at the values quoted in the caption of Table~\ref{table1}.}
\end{deluxetable}

As seen from Table~\ref{table2}, the impact parameter $u_\mathrm{min}$ is considerably smaller than the source size parameter $\rho$, implying the lens of MOA-2011-BLG-274 transited the source almost perfectly. For a perfect transit, i.e.~$u_{\mathrm{min}}=0$, we would have $A_{\mathrm{max}}=2\theta_{\mathrm{E}}/\theta_{\mathrm{S}}$ in the absence of limb darkening (Liebes, 1964). This would imply $A_{\mathrm{max}}=2\theta_{\mathrm{E}}/\theta_{\mathrm{S}}=2/\rho=190
$. As seen from Figure~\ref{fig3}, the effect of limb darkening is to effectively reduce the radius of the source slightly, and hence to increase $A_{\mathrm{max}}$ slightly. 

The angular Einstein radius is given by the relationship $\theta_{\mathrm{E}}=\theta_{\mathrm{S}}/{\rho}$ and is included in Table~\ref{table2}. As found by Choi et al.~it is considerably smaller than in typical microlensing events. As shown in Table~\ref{table2}, the measured value of the Einstein radius crossing time $t_{\mathrm{E}}$ = 3.26d is also anomalously small. Typically $t_{\mathrm{E}}\sim20$d (Sumi et al.~2011, Paczynski 1996) and  $\theta_{\mathrm{E}}\sim0.6$mas for a one-third-solar mass lens half way to the bulge. 

The above measurements independently suggest a small value for the mass of MOA-2011-BLG-274L, as the Einstein radius is proportional to the square root of the mass of the lens. But a small mass is not assured, as the above measurements do not fix the distance to the lens, $D_\mathrm{L}$, and the lens-mass depends on this as well. The magnitude of $D_\mathrm{L}$ is the subject of the following sections.
 
\section{Parallax I}

The importance of measuring the lens distance $D_\mathrm{L}$ may be seen if we examine the range of values that $M_\mathrm{L}$ takes if, as a worst case scenario, we eschew any knowledge of $D_\mathrm{L}$. In that case we may use Einstein's equation for the mass of a lens as a function of $\theta_\mathrm{E}$ and $D_\mathrm{L}$ as follows

\begin{equation}
~~~~~~~~~~~~~~~~~~{{M_\mathrm{L}}=\frac{{\mathrm{c}^2}{{\theta_\mathrm{E}}^2}}{\mathrm{4G}}\times{\frac{{D_\mathrm{S}}{D_\mathrm{L}}}{{D_\mathrm{S}}-{D_\mathrm{L}}}}}.~~~~~~~~~~~~~~~~~~
\end{equation}  

This is a monotonically increasing function of $D_\mathrm{L}$ with values $2.35M_\mathrm{J}$ at $D_\mathrm{L}$ = 2 kpc, $7.05M_\mathrm{J}$ at $D_\mathrm{L}$ = 4 kpc and $21.2M_\mathrm{J}$ at $D_\mathrm{L}$ = 6 kpc for  $\theta_{\rm{E}}=0.084$ mas. Thus, if MOA-2011-BLG-274L is in the disc it is a PMO, and if it is in the bulge it is a brown dwarf. This highlights the need to determine the lens distance. In principle, this can be achieved by the microlensing parallax method.   
      
The observed magnification in any microlensing event at any time depends on the angular separation between the lens and the source. At small separations and high magnifications we have $A\approx1/u_\mathrm{min}$. The magnification therefore varies slightly from point to point on the Earth's surface at any given time. Also, the time of peak magnification varies from point to point (Hardy \& Walker 1995, Gould et al.~2009). These effects are known as terrestrial parallax, and they depend on the distance to the lens. The closer the lens, the larger the effects. Both effects are undetectable unless the magnification varies unusually rapidly with time. Inspection of Figure~\ref{fig1} reveals rapid variation of the light curve of MOA-2011-BLG-274, and also good coverage of the light curve from different locations. It therefore appears ideal for investigation of the measurability of terrestrial parallax. In a previous detection of terrestrial parallax (Gould et al.~2009), $A_\mathrm{max}$ was 2,500 and $t_\mathrm{E}$ was 7 days.        

The related effect of orbital parallax, in which the Earth's non-rectilinear motion about the Sun is taken into account, can also be used to determine the distance to the lens in microlensing events (Gould 1992, Alcock et al.~1995). Orbital parallax is normally detectable in events with relatively long Einstein times, but $t_{\mathrm{E}}$ was anomalously short in MOA-2011-BLG-274. We may therefore anticipate that, if parallax is detectable in MOA-2011-BLG-274, it will have been caused predominantly by terrestrial parallax.    

The effects of parallax (both terrestrial and orbital) may be quantified by the two-dimensional vector $\pi_{\mathrm{\textbf{E}}}$ with east and north components $\pi_{\mathrm{\textbf{E},E}}$ and $\pi_{\mathrm{\textbf{E},N}}$ respectively (Gould 2004). The magnitude of the parallax vector $|\pi_{\mathrm{\textbf{E}}}|$ is defined to be $\mathrm{AU}/\tilde{r}_{\mathrm{E}}$ where $\tilde{r}_{\mathrm{E}}$ is the radius of the Einstein ring projected back to the observer's plane, and its direction is defined to be the direction of motion of the lens projected onto the observer's plane.

\begin{figure*}
\plotone{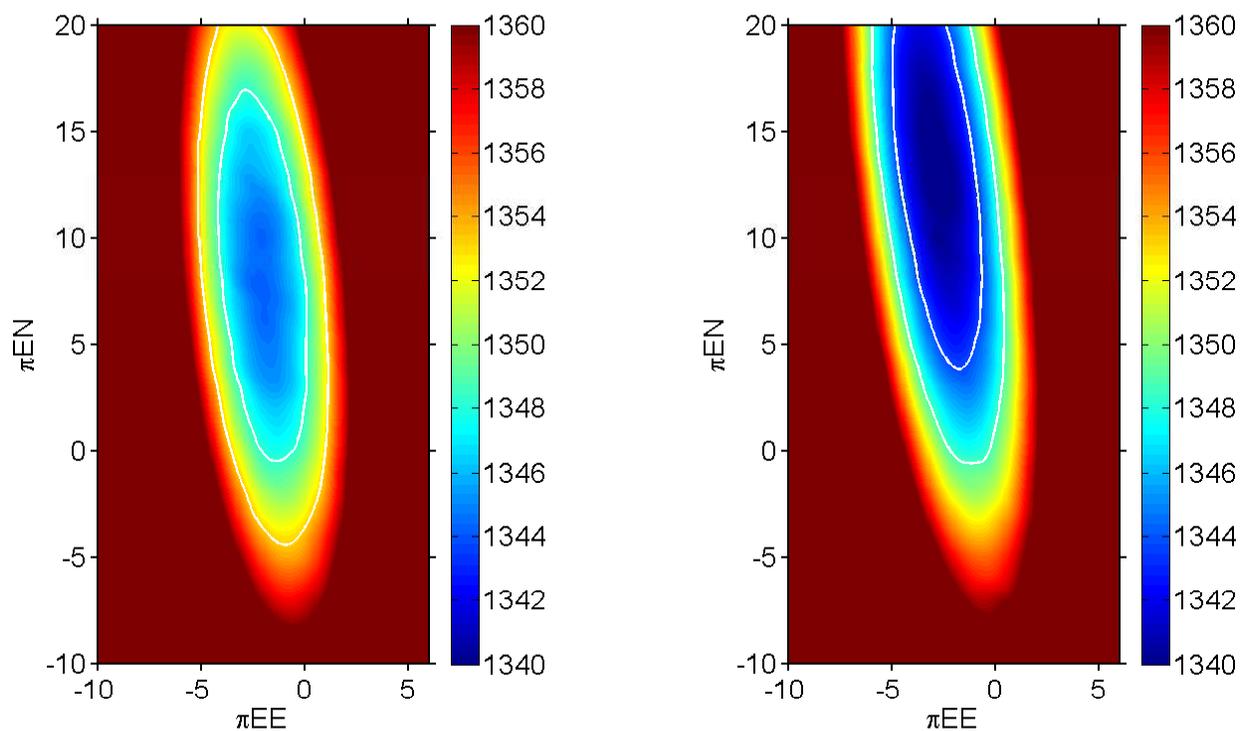}
\caption{$\chi^2$ maps of parallax for linear limb darkening at 5700K. The left and right panels show the best solutions with $u_\mathrm{min}$ negative and positive respectively. The inner and outer contours are at $\delta\chi^2$ = 4 and 9 respectively. The parameters for these and other solutions are given in Table~\ref{table3}.}
\label{fig5}
\end{figure*}

\begin{figure}
\plotone{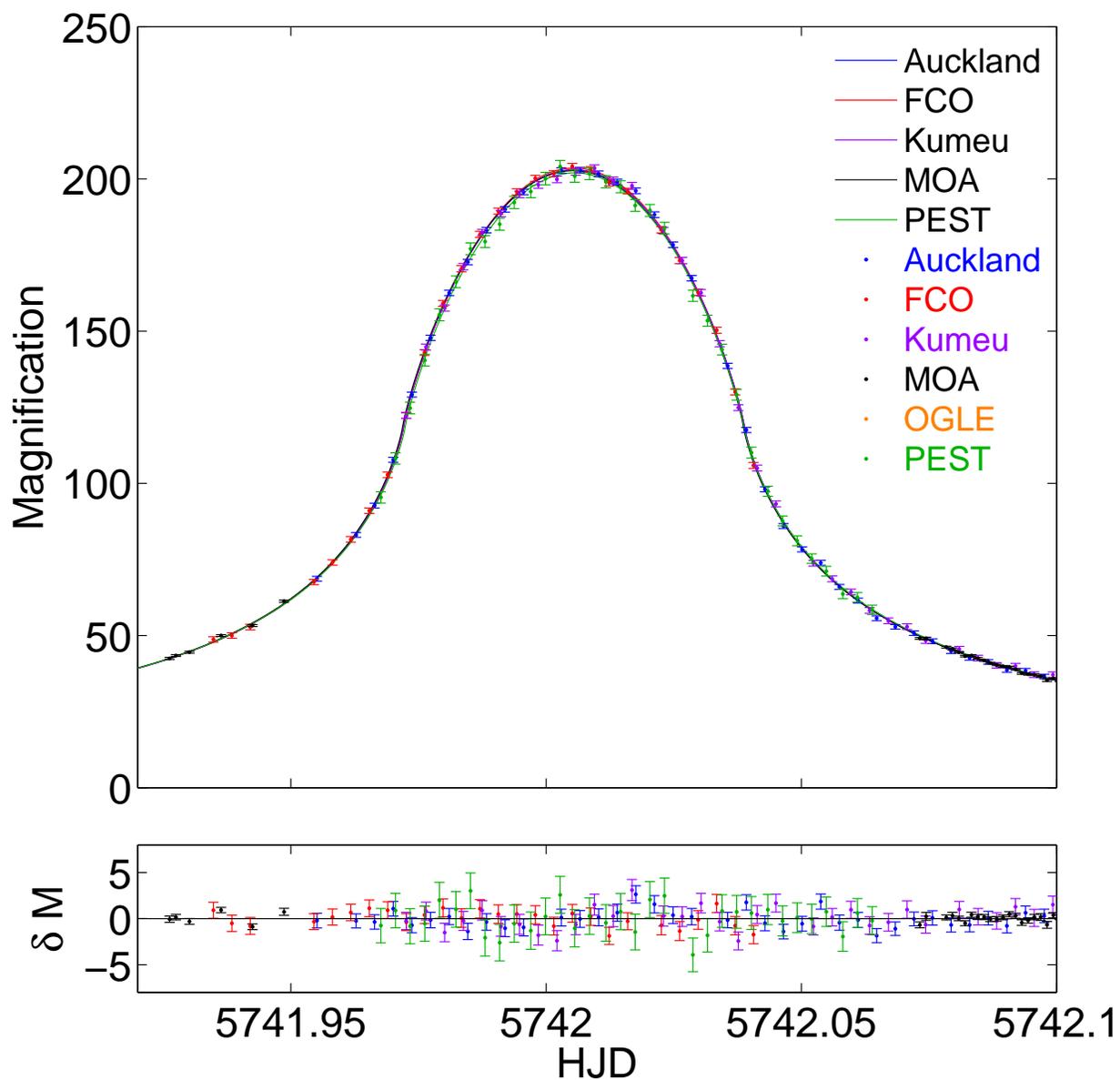}
\caption{Best fit light curves with parallax included, linear limb darkening, a source temperature of 5700K and positive impact parameter. Residuals are shown below. The small differences between these light curves and those shown in Figure~\ref{fig4} indicate the sensitivity required to measure parallax in events such as MOA-2011-BLG-274.}
\label{fig6}
\end{figure}

Best fits to the data for MOA-2011-BLG-274 were initially found over a coarse grid of values of the parallax plane with step sizes of 1 in both the easterly and northerly directions. Two minima were found for $(\pi_{\mathrm{\textbf{E},E}}$,$\pi_{\mathrm{\textbf{E},N}})$ at approximately (-3,+13) and (-2,+8) for positive and negative values of $u_{\mathrm{min}}$ respectively. The results for linear limb darkening and a source star temperature of 5700K are shown in Figure~\ref{fig5}. Fine grids were plotted over smaller regions of the parallax plane with step sizes of 0.1 for $\pi_{\mathrm{\textbf{E},E}}$ and $\pi_{\mathrm{\textbf{E},N}}$ for the best fitting model. These yielded the results in Table~\ref{table3}. The best fit light curves with parallax included are shown in Figure~\ref{fig6}. 

\begin{figure}
\plotone{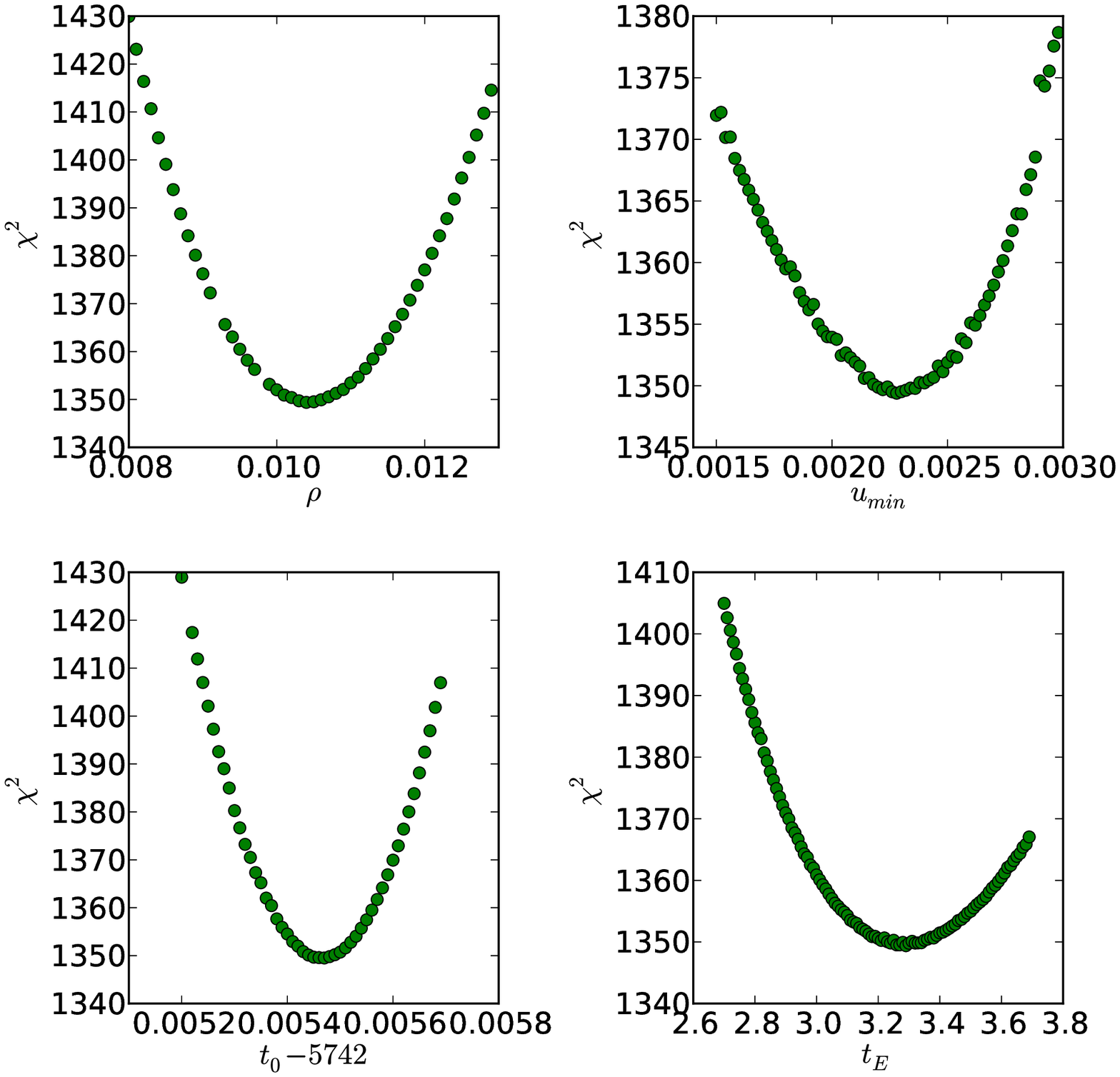}
\caption{Marginalization plots used to determine accuracies of measurements in Table~\ref{table3}. The uncertainty of each measurement was taken to be a quarter of the width of the fitted curve 2 sigma above the minium.}
\label{fig7}
\end{figure}

The values of parallax for all models in Table~\ref{table3} are an order of magnitude larger than found in most previous events. This implies an unusually small value for the Einstein radius projected back to the observer plane $\tilde{r}_{\mathrm{E}}$. MOA-2011-BLG-274 thus exhibited unusual behaviour on three counts, an unusually small angular Einstein radius $\theta_\mathrm{E}$, an unusually short Einstein radius crossing time $t_\mathrm{E}$, and an unusually large value of parallax. We note that the larger uncertainty of $\pi_{\mathrm{\textbf{E},N}}$ compared to that $\pi_{\mathrm{\textbf{E},E}}$ reflects the longer baseline between the telescopes in the east-west direction compared to the north-south direction. Uncertainties for $u_\mathrm{min}$,  $\rho$, $t_0$, and $t_{\mathrm{E}}$ were found by marginalising each parameter as shown in Figure~\ref{fig7}.      

\begin{deluxetable}{c c c c c c c c c c}
\tabletypesize{\scriptsize}
\tablecaption{Best fit parameters with parallax included.\label{table3}}
\tablehead{
\colhead{Limb darkening} & \colhead{$u_\mathrm{min}$} & \colhead{$\rho$} & \colhead{$t_0$ (HJD)} & \colhead{$t_{\mathrm{E}}$ (d)} & \colhead{$\theta_{\mathrm{E}}$ (mas)} & \colhead{$\mu$ (mas y$^{-1}$)} & \colhead{$\pi_{\mathrm{\textbf{E},E}}$} & \colhead{$\pi_{\mathrm{\textbf{E},N}}$} & \colhead{$\chi^2$}
}
\startdata
Linear (5700K) & +0.00236 & 0.0104 & 2545742.00574 & 3.30 & 0.084 & 9.29 & -2.8 & 13.2 & 1339.7\\
Linear (5500K) & +0.00228 & 0.0105 & 2545742.00574 & 3.27 & 0.083 & 9.26 & -3 & 13 & 1340.4\\
Linear (5900K) & +0.00238 & 0.0105 & 2545742.00573 & 3.26 & 0.083 & 9.29 & -3 & 13 & 1340.2\\
Square root & +0.00236 & 0.0104 & 2545742.00574 & 3.30 & 0.084 & 9.29 & -3 & 13 & 1340.1\\
Linear (5700K) & -0.00230 & 0.0105 & 2545742.00566 & 3.26 & 0.083 & 9.29 & -2 & 8 & 1344.3\\
Square root & -0.00238 & 0.0104 & 2545742.00566 & 3.29 & 0.084 & 9.32 & -2 & 8 & 1344.7\\
Uncertainty & $\pm$0.00014 & $\pm$0.0003 & $\pm$0.00003 & $\pm$0.44 & $\pm$0.011 & $\pm$1.90 & $\pm$1.2 & $\pm$4.6 & -\\ 
\enddata
\tablecomments{The $t_0$ values listed here differ from those listed in Table~\ref{table2}. The values in Table~\ref{table2} are the average peak times of all telescopes. The values in Table~\ref{table3} correspond to a hypothetical point at the centre of the Earth (as in Figure~\ref{fig9}).}
\end{deluxetable}

It is clear from Figure~\ref{fig5} that a detection of parallax of moderate significance only has been achieved, with a $\delta\chi^2$ improvement of 10 in comparison to the solution without parallax. This corresponds to $3\sigma$. 

In view of this, additional checks were made. Effects of differential refraction were searched for in the data obtained with the filterless PEST and FCO telescopes by examining the light curves of field stars of similar colour to the source star. The zenith angle for the PEST telescope decreased monotonically from approximately $55^{\circ}$ to approximately $33^{\circ}$ during the main night, but for the FCO telescope it was less than $20^{\circ}$ throughout the night. Comparison stars for the PEST telescope were found to drift by less than approximately $\pm5$ milli-magnitudes during the night, and for the FCO telescope the drift was less than $\pm10$ milli-magnitudes. No attempt was made to correct for these effects. Further checks are described in \S8 below.      
                
\section{Lens distance and mass}

\begin{figure}
\plotone{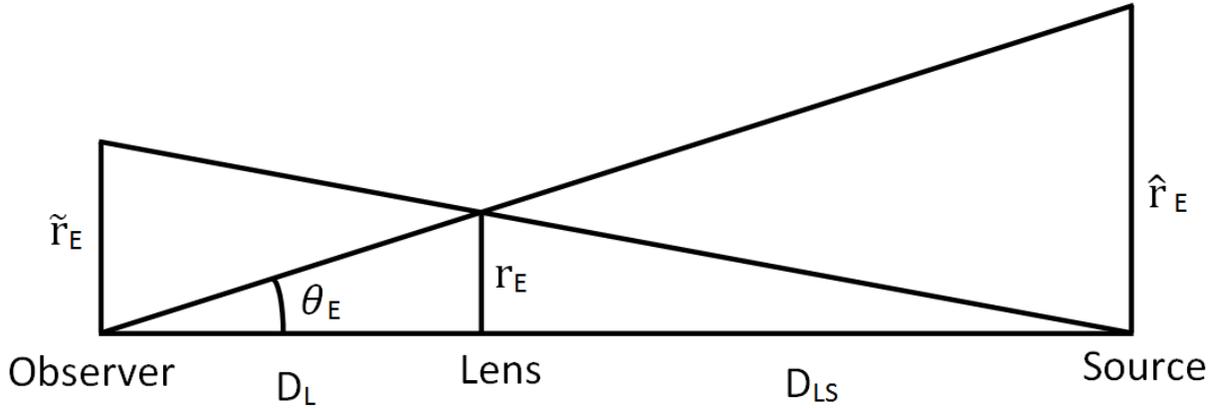}
\caption{Microlensing geometry showing the Einstein ring radius $r_{\mathrm{E}}$ projected onto the observer plane $\tilde{r}_{\mathrm{E}}$ and onto the source plane $\hat{r}_{\mathrm{E}}$. The distance to the source $D_{\mathrm{S}}$ is given by the sum of $D_{\mathrm{L}} + D_{\mathrm{LS}}$} 
\label{fig8}
\end{figure}

Figure~\ref{fig8}, which is adapted from Gould (2000), shows the Einstein radius  $r_{\mathrm{E}}$ projected simultaneously to both the observer plane $\tilde{r}_{\mathrm{E}}$ and the source plane $\hat{r}_{\mathrm{E}}$. Both of the projected radii were measured. We have 

\begin{equation}
\tilde{r}_{\mathrm{E}}=\frac{\mathrm{AU}}{|\pi_{\mathrm{\textbf{E}}}|}=\frac{\mathrm{AU}}{\sqrt{{\pi_{\mathrm{\textbf{E},N}}}^2+{\pi_{\mathrm{\textbf{E},E}}}^2}}= 0.074\pm0.025~\mathrm{AU} 
\end{equation}
and
\begin{equation}
\hat{r}_{\mathrm{E}}=\frac{r_{\mathrm{S}}}{\rho}=0.66\pm0.11~\mathrm{AU},
\end{equation} 
where the values of $\pi_{\mathrm{\textbf{E},E}}$, $\pi_{\mathrm{\textbf{E},N}}$, and $\rho$ are given in the first row of Table~\ref{table3}, and $r_{\mathrm{S}}$ is given in Equation (5). The confidence levels are 1 sigma.

It is apparent from Figure~\ref{fig8} that the values of $\tilde{r}_{\mathrm{E}}$, $\hat{r}_{\mathrm{E}}$, and $D_{\mathrm{S}}$ jointly determine the Einstein radius $r_{\mathrm{E}}$ and the distance to the lens $D_{\mathrm{L}}$. The figure leads to a `resistor' equation for the radii  
\begin{equation}
\frac{1}{r_{\mathrm{E}}} = \frac{1}{\tilde{r}_{\mathrm{E}}} + \frac{1}{\hat{r}_{\mathrm{E}}}, 
\end{equation}
which implies  
\begin{equation}
r_{\mathrm{E}}=0.067\pm0.020~\mathrm{AU}.
\end{equation}
Also, 
\begin{equation}
D_{\mathrm{L}}={\frac{r_{\mathrm{E}}}{\hat{r}_{\mathrm{E}}}}\times{D_{\mathrm{S}}}=0.80\pm0.25~\mathrm{kpc}.
\end{equation}
Finally, the mass of the lens $M_{\mathrm{L}}$ is given by Einstein's equation written in terms of $\tilde{r}_{\mathrm{E}}$ and $\theta_{\mathrm{E}}$  
\begin{equation}
~~~~~~~~~~~~~~~~~~~~{{M_\mathrm{L}}=\frac{{\mathrm{c}^2}{\tilde{r}_{\mathrm{E}}}{\theta_\mathrm{E}}}{\mathrm{4G}}= 0.80\pm0.30~M_\mathrm{J}}.~~~~~~~~~
\end{equation}

The lens of MOA-2011-BLG-274 thus has a mass similar to that of Jupiter with uncertainty arising mainly from uncertainty in the distance to the lens, $D_\mathrm{L}$, which itself arose from uncertainty in the measured value of parallax. 

\section{Trajectory} 
The trajectory of the lens of MOA-2011-BLG-274 projected to the observer plane may be determined. This assists to visualize the event, and also provides a further check on the above results.  
 
At any given time there is a point on the observer plane where the angular separation between the source and the lens will be zero. As the lens and source move in the sky, this point sweeps across the observer plane.  An observer on this line would measure $u_\mathrm{min}$ = 0. Using the parallax vector it is possible to calculate where this line passes in relation to the Earth.

The best fit for MOA-2011-BLG-274 has $u_\mathrm{min}$ = 0.00236. Projecting this separation to the observer plane gives ${u_\mathrm{min}}\times{\tilde{r}_\mathrm{E}}=0.00236\times0.074\mathrm{AU}=4.1\pm1.4$ Earth radii. The point of maximum magnification therefore swept past Earth a distance of 4.1 Earth radii away. Its direction was $\mathrm{tan}^{-1}(-2.8/13.2) = 12^{\circ}$ west of north as determined by the components of the parallax vector. The maximum traversed the projected Einstein radius $\tilde{r}_\mathrm{E}$ in the Einstein crossing time $t_{\mathrm{E}}$, so its speed was 39 kms$^{-1}$. A speed of this magnitude would arise from the known velocity dispersion of stars in the galactic disc. 

\begin{figure}
\plotone{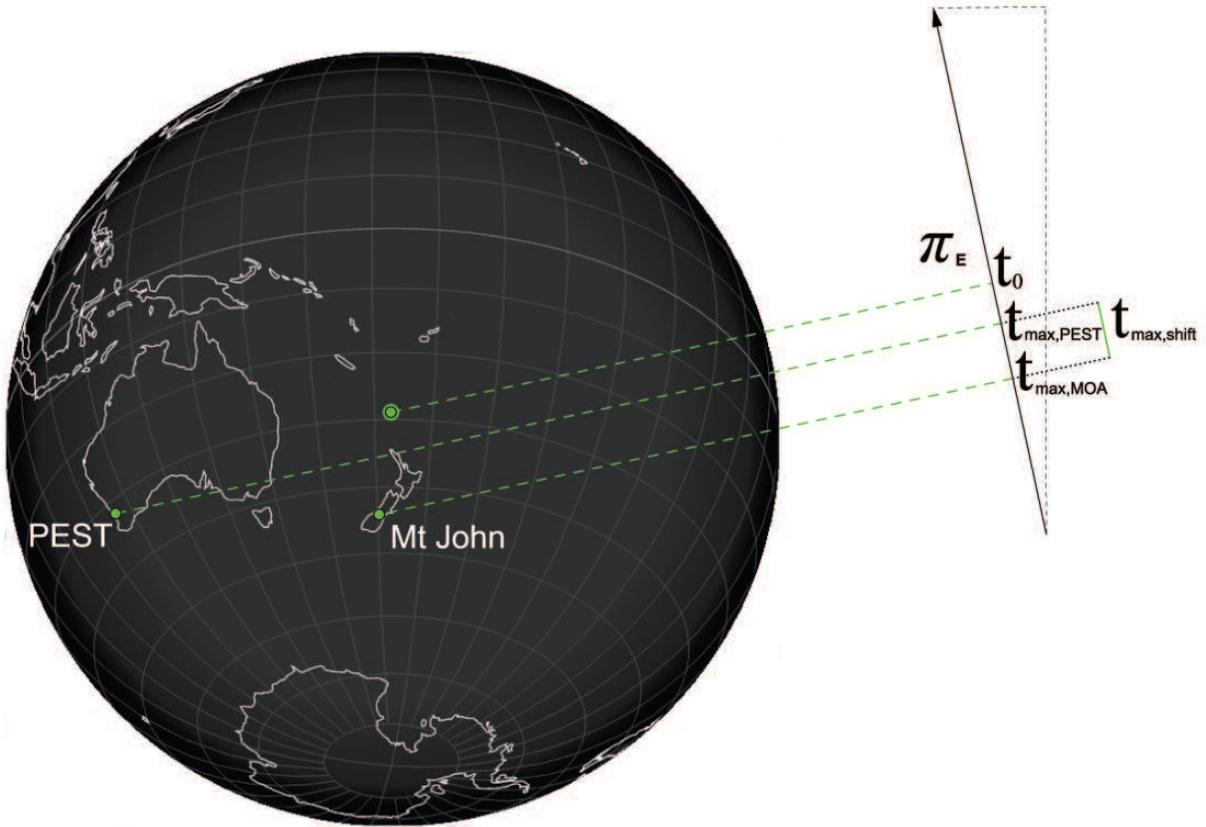}
\caption{Trajectory of MOA-2011-BLG-274 for the solution with impact parameter $u_{\mathrm{min}}=+0.00238$. The impact parameter has been reduced from its true value of 4.1 Earth radii in the figure to save space. A second possible trajectory with $u_{\mathrm{min}}=-0.00230$ passes the Earth at a similar distance on the westward side. A comparable diagram appears in Gould et al.~(2009).}
\label{fig9}
\end{figure}

The times of maximum magnification recorded by each telescope according to the best fit to the data (i.e.~the first model in Table~\ref{table3}) also assist to visualize MOA-2011-BLG-274. They were determined from the light curves for each telescope shown in Figure~\ref{fig6} and they are listed in Table~\ref{table4}. They are clearly consistent with an event travelling northwards. The projected distance between Mt John and the Auckland observatories along the above trajectory at $12^{\circ}$ west of north is 690 km, so at a speed of 39 km/s we anticipate an 18 second difference between the observed peak magnifications, in agreement with the times in Table~\ref{table4}. The peak magnification observed at Farm Cove was higher than that observed at Perth in the model with $u_{\mathrm{min}>0}$, indicating that the track lay to the East of Australasia in this model. A similar conclusion follows from the peak magnifications observed at Auckland and Mt John. An overall visualization of the trajectory of MOA-2011-BLG-274 is shown in Figure~\ref{fig9}. 

\begin{deluxetable}{l c c}
\tablewidth{0pt}
\tablecaption{Times of peak magnification for the individual telescopes.\label{table4}}
\tablehead{
\colhead{Telescope} & \colhead{$t_\mathrm{max} (\textrm{HJD})$} & \colhead{$t_\mathrm{max}$-$t_\mathrm{max,MOA}$ (\textrm{seconds})}
}
\startdata
PEST &	5742.005594 &	27.2\\
Kumeu & 5742.005490 & 18.2\\
Auckland &	5742.005484 &	17.7\\
Farm Cove &	5742.005483 &	17.6\\
MOA &	5742.005279 &	-\\
\enddata
\tablecomments{These are determined by the best overall fit to the data for all telescopes shown in Figure~\ref{fig6}. The peak times were taken to be the mid-times when the magnification was 185 for the given site on the ascending and descending branches of the event.}
\end{deluxetable}

\section{Parallax II}
The analysis in \S5 includes all the available information on parallax in MOA-2011-BLG-247 in compact form. However, it is perhaps not as transparent as possible. For example, it does not isolate the terrestrial component of parallax from the orbital component. Also, it is difficult to see with the naked eye any improvement of the light curve with parallax (Figure~\ref{fig6}) over that without (Figure~\ref{fig4}).  

In the present event we expect orbital parallax to be undetectably small as the event occurred at the end of June when the Earth's orbit is nearly perpendicular to the line of sight to the Galactic bulge, and because the Einstein crossing time was short. We therefore expect terrestrial parallax to dominate. The clearest manifestation of the latter is the different times of peak magnification it causes for telescopes at different locations on the Earth's surface. An analysis that focuses on these times may therefore be helpful. 

\begin{figure}
\plotone{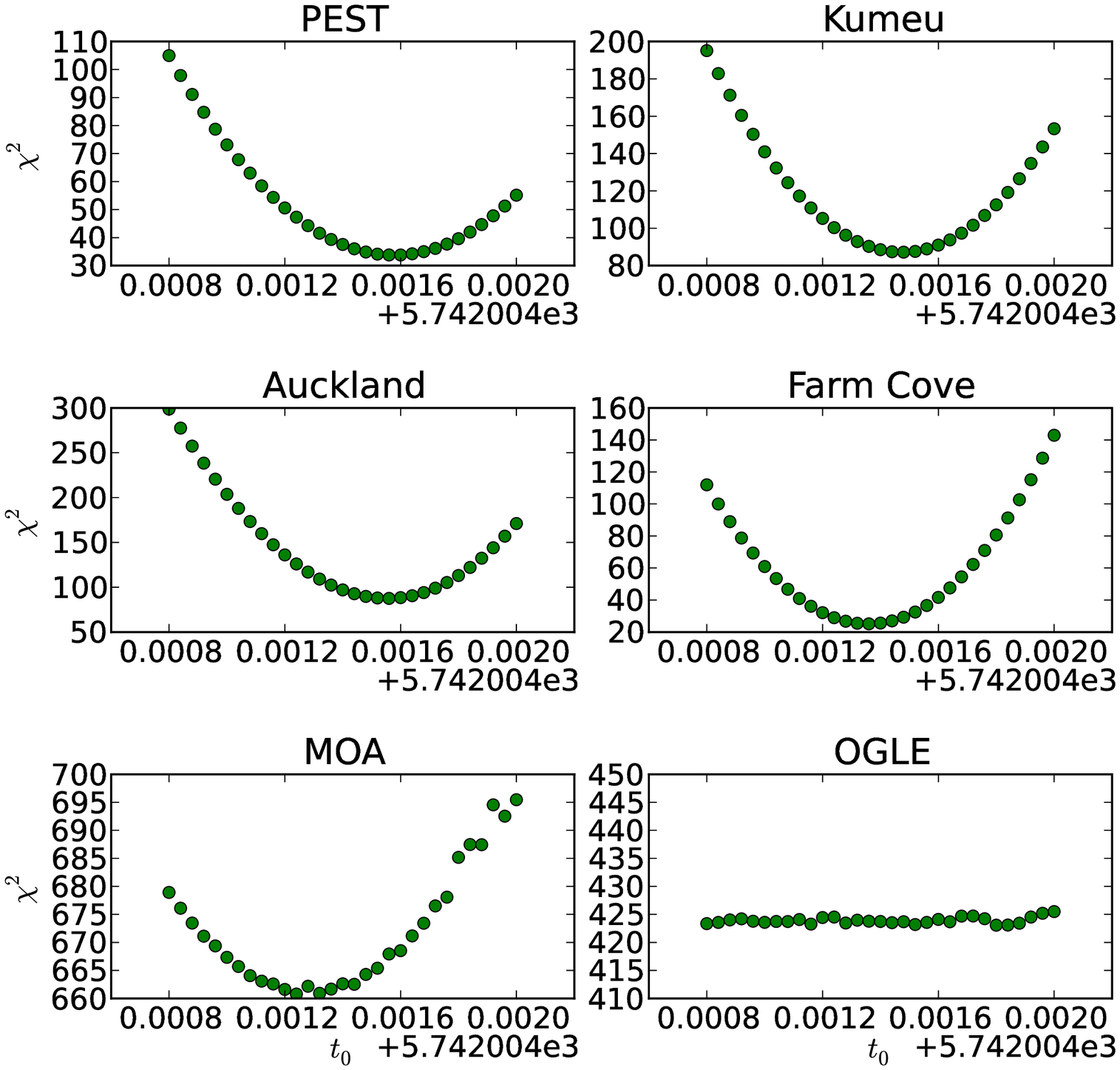}
\caption{Individual telescope timing plots as described in the text with parallax excluded. No information is available from the OGLE telescope as it was in daylight at the peak of the event.}
\label{fig10}
\end{figure}

With this in mind, the data for each telescope were individually fitted to single-lens, finite-source, linearly limb-darkened, non-parallax light curves for a sequence of values of $t_\mathrm{0}$. In this procedure $u_\mathrm{min}$, $\rho$, and $t_{\mathrm{E}}$ were allowed to vary. The results are shown in Figure~\ref{fig10}. These plots enable the effect of parallax to be seen by the naked eye. They were used to extract the peak times recorded by each telescope together with their uncertainties, as given in Table~\ref{table5}.   

\begin{deluxetable}{l c c}
\tablewidth{0pt}
\tablecaption{Peak magnification times and uncertainties recorded by individual telescopes.\label{table5}}
\tablehead{
\colhead{Telescope} & \colhead{$t_\mathrm{max} (\textrm{HJD})$} & \colhead{$t_\mathrm{max}$-$t_\mathrm{max,MOA}$ (\textrm{seconds})}
}
\startdata
PEST &	$5742.00558\pm0.00009$ &	$26\pm8$\\
Kumeu & $5742.00547\pm0.00006$ & $16\pm5$\\
Auckland &	$5742.00554\pm0.00005$ &	$22\pm4$\\
Farm Cove &	$5742.00536\pm0.00006$ &	$7\pm5$\\
MOA &	$5742.00528\pm0.00012$ &	$0\pm10$\\
\enddata
\tablecomments{These are determined by the timing plots in Figure~\ref{fig10}. The right hand column shows the same data converted to seconds with the MOA time taken as a nominal zeropoint.}
\end{deluxetable}

Comparison of the entries in Tables~\ref{table4} and~\ref{table5} shows that the times recorded by the individual observatories agree closely with those determined by the combined analysis in \S5 of all data from all observatories with the exception of the time recorded by the Farm Cove observatory where there is a $2\sigma$ discrepancy. We assume this occurred as a statistical fluctuation. 

The above highlights the accuracy of the timing needed to carry out a successful measurement of terrestrial parallax in events like MOA-2011-BLG-274. The required accuracy is of order a few seconds for the peak of the light curve for each telescope. This is undoubtedly demanding, but we note that the analysis is expected to be relatively immune to small uncertainties in limb darkening, sky transmittance and CCD spectral response of each telescope, and also to small uncertainties in the best values of $u_{\mathrm{min}}$, $\rho$, and $t_{\mathrm{E}}$ used in the fitting procedure. This follows because only the axis of symmetry of each light curve was extracted from the data, and small errors in the above quantities should not affect this appreciably, especially as each telescope observed the ascending and descending branches of the light curve almost symmetrically. 

Suppose, as discussed in \S5, the lens for the event was actually a brown dwarf at a distance $\geq6$ kpc. Figure~\ref{fig8} then predicts a velocity of the lens in the observer plane $\geq1050$ $\rm{kms^{-1}}$. Whilst such a velocity may be possible with high velocity stars in the bulge, it would imply timing differences between the various telescopes in Table~\ref{table5} $\leq1$s which appears unlikely.   

Checks were made of the sampling procedure used to determine the light curve. The most important data were obtained in a four hour time span when the light curve was sampled in 300 sec exposures (Auckland, FCO, Kumeu), 240 sec exposures (PEST), 100 sec exposures (OGLE) or 60 sec exposures (MOA). The luminosity changed appreciably during these exposures, but it was implicitly assumed that the average magnification during an exposure equalled the magnification at the mid-exposure time. 

\begin{figure}
\plotone{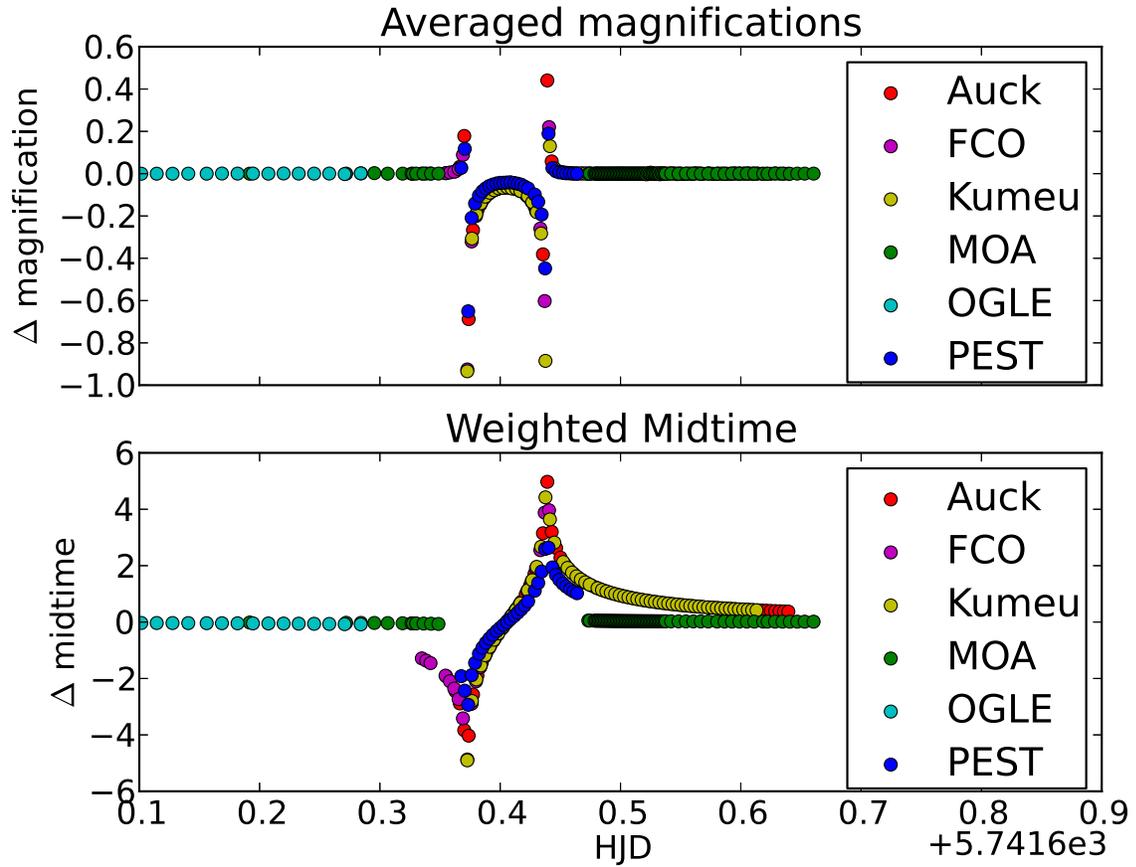}
\caption{Effects of finite exposure times. The upper panel shows the difference between the average magnification in an exposure and the magnification at mid-exposure. The lower panel shows the difference (in secs) between the average photon arrival time in an exposure and the mid-exposure time.}  
\label{fig11}
\end{figure}

Figure~\ref{fig11}, which is based on the best fit to the data, quantifies this effect. The upper panel shows the difference between the average and mid-exposure magnifications by telescope, and the lower panel shows the difference between the average photon arrival time and the mid-exposure time. Both plots highlight the points of inflexion on the lightcurve where the effects are greatest. If the portion of the light curve immediately before and after a point of inflexion is non-uniformly monitored by a telescope, then an apparent displacement $\sim1-2$ sec could result in the measured peak time. We expect this is approximately the magnitude of this effect in the present event, and suggest it could be advisable in future events to limit exposures to 200 seconds.   

\section{Host star}
The preceding discussion indicates that the lens in MOA-2011-BLG-274L may be an isolated PMO or a sub brown dwarf. In principle, it may also be a planet orbiting a host star at such a large distance that the host did not affect the microlensing light curve appreciably.  

To test this possibility simulations were conducted with a star of mass 300 times the mass of the planet, i.e.~approximately one-third solar mass or the most likely value, located at various distances out to 1000 Einstein radii from the planet, i.e.~out to approximately 70 AU from the planet.

Our simulations, and those in the following section, were conducted using the magnification map procedure described in Abe et al (2013) and the on-line reduction of the images from the MOA telescope. The planet was placed at the origin in the lens plane, and its possible host star with a star:planet mass ratio of $q_2$ was placed on the $y$-axis at $y_2$. Both $q_2$ and $y_2$ were assumed to be large in the sense that ${q_2}\gg1$ and ${y_2}\gg1$. Thus the host star was assumed to be much heavier than the planet, and its separation from the planet was assumed to be much larger than the Einstein radius of the planet. 



Under these conditions, the point of maximum magnification in the source plane is no longer directly behind the lens. It is shifted by a factor of $q_2/y_2$ Einstein radii towards the larger second body (Dominik 1999). Also, it becomes an astroid shaped four-fold caustic, which is transited by the source in the best fit.

\begin{figure}
  \plotone{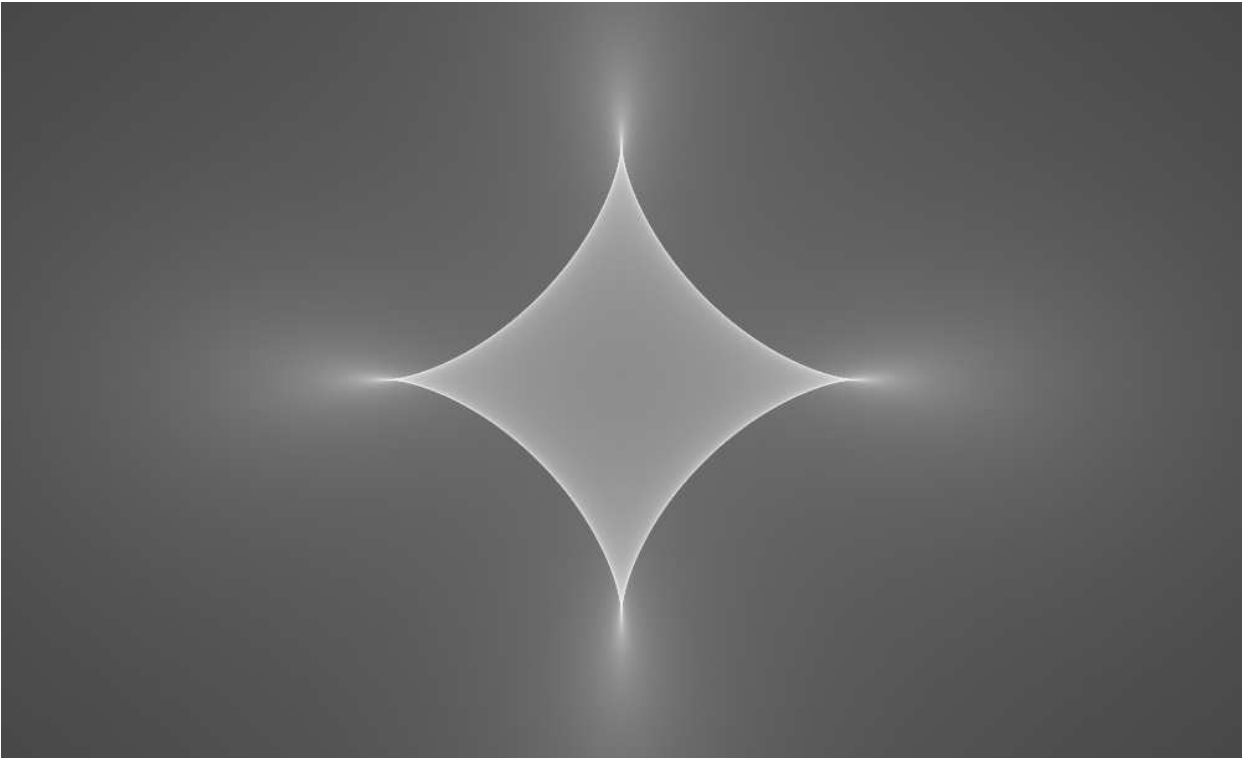}
  \caption{Magnification map for a planet with a distant host star with $q_2$ = 300 and $y_2$ = 400. The diameter of the caustic is $\approx1.3\times$ the diameter of the source star, and its effect on the light curve is detectable ($\delta\chi^2\approx150$).}
  \label{fig12}
\end{figure}

Figure~\ref{fig12} shows a typical magnification map, a four-fold caustic, when the host star is detectable ($q_2=300$ and $y_2=400$). The horizontal and vertical diameters of the caustic are similar to the diameter of the source star, and the effect of the caustic on the light curve is detectable. For host stars at greater distances with $y_2\geq500$ the size of the caustic diminishes and it fits completely inside the source star. Its effect on the light curve is then undetectable. Similar behaviour is well-known in the case of lensing by an individual star on the outskirts of a galaxy (Chang and Refsdal, 1979). 

The results are shown in Table~\ref{table6}. A host star with $m_2 = 300$ or approximately 0.2 solar mass can be excluded out to 400 Einstein radii, i.e.~out to approximately 27 AU or approximately the orbit radius of Neptune. Such a star would have caused a detectable deviation on the light curve. Simulations with a star at larger distances showed little change in  $\chi^2$ over the single lens fit. Simulations were also conducted for host star masses of 0.08, 0.4 and 0.8 solar masses. It was found they could be excluded out to 280, 600 and 900 Einstein radii, or approximately 19, 40 and 60 AU respectively.

A search for a host star at a separation $\geq27$ AU could be carried out with the Hubble Space Telescope. At a distance of 0.8 kpc a separation $\geq27$ AU corresponds to an angular separation $\geq.03$ arcsec. A host star would therefore be detectable as a spatially resolved object close to the source star, or as additional flux exceeding the measured flux of the source star given in $\S3$. Absence of such a star would confirm the interpretation of the event as an isolated PMO.     

\begin{deluxetable}{cccccccc}
\tablewidth{0pt}
\tablecaption{Host Star.\label{table6}}
\tablehead{
\colhead{Mass} & \colhead{Distance} & \colhead{$u_\mathrm{min}$} & \colhead{$\psi$} & \colhead{$\rho$} & \colhead{$t_0$} & \colhead{$t_\mathrm{E}$} & \colhead{$\chi^2$}\\
\colhead{($q_2$)} & \colhead{($y_2$)} & & \colhead{(rad)} &  & \colhead{(HJD)}
}
\startdata
300 & 300 & 0.00009 & 0.4 & 0.0112 & 5742.00543 & 2.96 & 2487.64\\
300 & 400 & -0.00011 & 0.25 & 0.0122 & 5742.00546 & 2.78 & 1507.45\\
300 & 500 & 0.00163 & 0.05 & 0.0112 & 5742.00545 & 3.05 & 1400.70\\
300 & 700 & 0.00233 & 0.25 & 0.0112 & 5742.00544 & 3.06 & 1363.96\\
300 & 800 & 0.00245 & 0.65 & 0.0112 & 5742.00543 & 3.06 & 1360.15\\
300 & 900 & 0.00255 & 0.55 & 0.0112 & 5742.00543 & 3.065 & 1360.52\\
300 & 1000 & 0.00255 & 0.492 & 0.0112 & 5742.00543 & 3.065 & 1361.70\\
\enddata
\end{deluxetable}

\section{Satellites}
Conceivably, isolated PMOs may have satellites orbiting them, sometimes referred to as `exomoons'. Searches by microlensing have been carried out for such objects, e.g.~(Bennett et al., 2013). Choi et al.~(2012) already reported a search for satellites to MOA-2011-BLG-274L. Using our magnification map procedure we were able to confirm the results reported by Choi et al. We did not find any evidence for the presence of satellites, and we were able to exclude them in the regions found by Choi et al. 

In fact, we derived slightly larger exclusion regions than those reported by Choi et al. We found a small but finite exclusion zone for satellites with a satellite:host mass ratio of $q=10^{-4}$ whereas Choi et al.~reported none. At $q=10^{-3}$ Choi et al.~reported approximately 50\% exclusion at $d=0.7~r_\mathrm{E}$ whereas we found approximately 90\%. Jupiter's largest moon, Ganymede, has a mass ratio of $7\times10^{-5}$. Unfortunately, this lies just beyond the level of detectability achieved in MOA-2011-BLG-274.   

\section{Discussion and future observations}
This is the first time that a direct measurement of the mass of an isolated PMO has been attempted. The masses of PMOs found previously in star forming regions were derived from theoretical cooling curves, and the masses of PMOs found previously by microlensing were deduced using statistical arguments.   

MOA-2011-BLG-274 was unusual because the lens transited the source, and because both the Einstein angular radius and the Einstein radius crossing time were consistent with values expected for a PMO. Our attempt to measure terrestrial parallax in the event yielded a result at the $3\sigma$ level of confidence. This led to mass and distance values for the lens of $0.80\pm$ 0.30 $M_{\mathrm{J}}$ and $0.80\pm0.25$ kpc respectively. The former value is consistent with expectation based on the results of Sumi et al.~(2011).  

Most of the crucial observations of MOA-2011-BLG-274 were carried out with `backyard' telescopes with apertures $\sim0.35$m located in New Zealand or Australia. If future events were observed with 1m-class  or larger telescopes distributed over a larger portion of the Earth's surface, then it appears possible that highly significant results on PMOs would emerge. 

However, this could necessitate a change of current observing strategies. The critical observations by backyard telescopes of MOA-2011-BLG-274 were made continuously. If future events were observed with larger telescopes, these observations would also need to be made continuously, or nearly continuously, in order to realise a significant improvement in accuracy over the backyard observations reported here. 

At present this is not a strategy that is generally employed, but it could be implemented by groups such as the Las Cumbres Observatory Global Telescope Network (LCOGT)\footnote{https://lcogt.net}. Certainly the geographical location of the LCOGT telescopes is ideal, with three sites in the southern hemisphere and two in the north.

A second problem is the difficulty of triggering efficiently on scarce, rapid events. The scarcity of transiting PMO events may be seen as follows. The MOA and OGLE collaborations observe a few events per year in which normal stellar lenses transit a main sequence source star. If PMOs are approximately twice as abundant as stars, as the results of Sumi et al.~(2011) indicate, we may expect twice as many events per year in which a PMO transits a source. This suggests a few suitable events per year in the fields presently monitored by MOA and OGLE that are close enough ($D_L < 3$ kpc) to permit a parallax measurement. A similarly low rate was estimated recently by Gould and Yee (2013). With this low rate, a network such as LCOGT would need to trigger efficiently on suitable events. The fact that this was done successfully in the case of MOA-2011-BLG-274 demonstrates it is possible.
 
Another possibility would be for a network such as the Korean Microlensing Telescope Network (KMTNet)\footnote{http://www.kasi.re.kr/english/project/KMTNet.aspx} to monitor two (say) fields alternately rather than the greater number they presently plan to observe (Henderson et al,~2014). This would yield exquisite data on fewer events, but it should include a sample of high-quality transiting PMO events. This option would also enjoy sensitivity to PMOs of lower masses than those reported by Sumi et al. It is now known that bound terrestrial planets outnumber bound giant planets by a factor of a few (Cassan et al.~2012). If a similar distinction applies to isolated PMOs, then the KMTNet may be able to detect PMOs down to super-Earth mass, assuming that their strategy could be modified.                

\section{Conclusion}

We re-analysed the short time-scale, high-magnification microlensing event MOA-2011-BLG-274 that was monitored by 0.3m-class and larger telescopes. We used re-reduced photometry and confirmed the possible interpretation of the event by Choi et al.~(2012) in terms of an isolated PMO. We attempted to carry out a measurement of terrestrial parallax in the event and obtained a result at the $3\sigma$ level of confidence. This corresponded to a mass of $0.80\pm$ 0.30 $M_{\mathrm{J}}$ for the lens of the event. 

We proposed observational strategies employing high-cadence observations with 1m-class telescopes to enable higher quality measurements to be made in future events. These observations would require a concerted effort by the microlensing community but they would be sensitive to PMOs from Jupiter mass to terrestrial mass.     

\section*{Acknowledgements}
We would like to thank the referee for helpful comments and suggestions. IB and PY acknowledge support by the Marsden Fund of New Zealand, FA and YM by JSPS grants 22403003, 23340064 and 23654082, ZD by the Fondation de Coop\'{e}ration Scientifique Paris-Saclay, AG by NSF grant AST 1103471 and NASA grant NNX12AB99G, DH by a Czech Science Foundation grant GACR P209/10/1318, CH by the Creative Research Initiative Program (2009-0081561) of National Research Foundation of Korea, JS by (FP7/2007-2013) / ERC grant agreement no.~246678 awarded by the European Research Council under the European Community's Seventh Framework Programme to the OGLE project and NR by a Royal Society of New Zealand Rutherford Discovery Fellowship.

\bibliographystyle{plainnat}

\appendix

\label{lastpage}

\end{document}